\documentstyle[epsfig]{elsart}

\def\lapp{\;\raisebox{-.5ex}{$\stackrel{<}{\scriptstyle\sim}$}\;}
\begin{document}

\hoffset=-0.3truecm
\vglue1.5truecm
\centerline{{\bf IN-MEDIUM $\omega$-MESON BROADENING AND S-WAVE}}
\smallskip
\centerline{{\bf PION ANNIHILATION INTO $e^+e^-$ PAIRS}}
\vskip 1true cm
\centerline{{\bf Gyuri Wolf}}
\vskip 0.3 true cm
\centerline{{\it CRIP, H-1525 Budapest, Hungary}}
\vskip 0.3 true cm
\centerline{{\bf Bengt Friman}}
\vskip 0.3 true cm
\centerline{{\it GSI, Planckstrasse 1}}
\centerline{{\it D-64291 Darmstadt, Germany}}
\centerline{{\it Institut f\"ur Kernphysik, TH Darmstadt}}
\centerline{{\it D-64289 Darmstadt, Germany}}
\vskip 0.3 truecm
\centerline{and}
\vskip 0.3 truecm
\centerline{{\bf Madeleine Soyeur}}
\vskip 0.3 true cm
\centerline{{\it D\'epartement d'Astrophysique, de Physique des Particules,}}
\centerline{{\it de Physique Nucl\'eaire et de l'Instrumentation Associ\'ee}}
\centerline{{\it Service de Physique Nucl\'eaire}}
\centerline{{\it Commissariat \`a l'Energie Atomique/Saclay}}
\centerline{{\it F-91191 Gif-sur-Yvette Cedex, France}}
\vskip 0.5 truecm
\noindent
\centerline {{\bf Abstract}}\par
\vskip 0.5truecm
\parshape=1 1truecm 14.0truecm \noindent
An $\omega$-meson in motion with respect to a nuclear medium can
couple to a $\sigma$-meson through a particle-hole excitation.  This
coupling is large. We investigate its consequences for the width of
$\omega$-mesons in matter and for the s-wave annihilation of pions
into lepton pairs
which can take place
in relativistic heavy ion collisions.  We find that
the two pion decay of $\omega$-mesons, resulting from the
$\omega-\sigma$ transition and the subsequent $2\pi$ decay of the
$\sigma$-meson, leads to a substantial broadening of $\omega$-mesons
in matter and possibly to an observable effect in experiments
measuring the $e^+e^-$ decay of vector mesons produced in nuclei and
in relativistic heavy-ion collisions. The inverse process, the s-wave 
annihilation of pions into $\omega$-mesons decaying into $e^+e^-$ pairs, 
has in general a much smaller
cross section than the corresponding p-wave annihilation through
$\rho$-mesons and is expected to contribute rather
little to the total $e^+e^-$ pair production in relativistic heavy ion 
collisions.
\vfill\eject
\noindent {\bf 1. Introduction}\par
\vskip 0.4 truecm
\noindent
In general, a meson which interacts with a nuclear medium 
does not have well-defined quantum numbers as in free space.
Both continuous space-time symmetries (such as Lorentz invariance)
and discrete symmetries (parity, charge conjugation, G-parity) can be
broken. For example, the motion of a meson which propagates 
with respect to the medium with a 3-momentum 
$\vec q \not = 0$ introduces a
spatial direction which breaks rotational invariance and allows for
the mixing of meson states of different total angular momenta. To lowest
order, this mixing is generated by the coupling of the mesons to
particle-hole excitations. We restrict ourselves to symmetric nuclear
matter, a situation in which isospin is conserved.
\par
Of particular interest is the mixing of vector and scalar mesons
in nuclear matter. In the isovector channel, the $\rho$-exchange
is a major part of the short-range
particle-hole interaction which inhibits pion condensation \cite{Brown-Weise}.
In the isoscalar channel, the coupling
of the $\sigma$ and $\omega$ fields through particle-hole
excitations is extremely large and is therefore expected to
strongly affect 
the propagation of $\omega$-mesons in nuclear matter \cite{Celenza,Jean}.
\par
In view of the role that vector mesons play both in hadronic physics and
in nuclear dynamics,
the propagation of $\rho$- and $\omega$-mesons in matter 
is at present the subject
of many theoretical investigations. These studies have been largely
motivated by the suggestion that in-medium
vector meson masses
could be order parameters for chiral symmetry restoration at
large baryon density \cite{Brown-Rho}.
In this case, the masses of the $\rho$- and $\omega$-mesons
would decrease
with increasing density as a consequence of their close relation to the
quark condensate. To test this idea, measurements of the
$e^+e^-$
spectrum of vector mesons produced in nuclei and in
heavy ion collisions are planned at TJNAF \cite{CEBAF} and GSI \cite{HADES}.
The kinematics is chosen so that at least $\rho$-mesons
decay inside the nuclear medium.
We recall that the $\rho$(770) and $\omega$(782) have similar masses
but very different widths,
$\Gamma_\rho\,=\,$ (151.2 $\pm$ 1.2) MeV and
$\Gamma_\omega\,=\,$ (8.43 $\pm$ 0.10) MeV, corresponding to
propagation lengths of c$\tau_\rho$ = 1.3 fm
and c$\tau_\omega$ = 23.4 fm in free space \cite{PDG}.
The $e^+e^-$
spectrum will reflect the spectral function of vector
meson-like excitations in matter. At sufficiently low density
and temperature,
it is expected that these excitations will behave as quasi-particles
which can be characterized by a mass and a width.\par
The in-medium width of $\rho$- and $\omega$-mesons plays a major role
in the observability of their in-medium mass.
The in-medium widths of vector mesons determine their mean free path
in matter and the probability that they decay inside the nuclear
target (or in the interaction region for heavy ion collisions)
after they are produced. In this respect, the medium broadening
of the $\omega$-meson is particularly important because it could
make its
lifetime comparable to the size of the nuclear system in which
it is produced. 
\par
The dynamical broadening of $\omega$-mesons in matter
can come from elastic and inelastic processes \cite{Klingl}.
Collisional broadening is expected 
in particular from the $\omega$ N $\rightarrow$
$\pi$ N reaction which has a fairly large cross section 
and can cause a significant absorption of $\omega$-mesons into the pion 
channel. 
By applying detailed balance we find, 
using the measured $\pi$ N $\rightarrow$ 
$\omega$ N cross section \cite{Karami}, a contribution to the $\omega$ 
width in matter of $\approx 10$ MeV at normal nuclear density \cite{Friman98}.
The  $\omega$ N $\rightarrow$
$\rho$ N reaction can also play a role and lead to the absorption 
of $\omega$-mesons into the p-wave two-pion channel.
In this paper, we work out the additional width of $\omega$-mesons
in matter arising from the large $\omega \, \sigma$
mixing in matter. The $\sigma$ meson is an effective degree of freedom
which parametrizes the propagation of two pions in s-wave.
The coupling of the $\omega$ to the $\sigma$ in matter
opens therefore a new two pion decay channel for the
$\omega$-meson.
We show that
the associated broadening of the $\omega$-meson is large
and increases with baryon density as well as with the momentum
of the $\omega$-meson. Conversely, the in-medium coupling of
the $\sigma$ to the $\omega$ makes it possible for two pions in s-wave
to annihilate into a lepton pair through an intermediate
$\omega$-meson \cite{Weldon,Kunihiro}. We discuss the consequences of this
new annihilation channel for the production of
$e^+e^-$ pairs in relativistic heavy ion collisions.
\par
In Section 2, we compute the $\omega \sigma$ polarization
at finite density and temperature in the one-loop approximation.
The $\sigma \pi \pi$ coupling constant and the $\sigma$
mass are fitted to the
$\pi \pi$ s-wave scattering phase shifts in Section 3.
Our numerical results for the in-medium $\omega$
broadening and s-wave pion annihilation into
$e^+e^-$ pairs are discussed in Section 4 and
we conclude with a brief discussion in Section 5.
\par
\vskip 0.4 truecm
\noindent
{\bf 2. Calculation of the $\omega$$\sigma$ polarization}\par
\vskip 0.4 truecm
\noindent
The polarization operator describing the mixing of $\omega$-
and $\sigma$-mesons to
lowest order in matter is shown in Fig. 1.
At sufficiently low temperatures, it is dominated by
intermediate nucleon-hole states. 
The two processes discussed in this paper, the broadening of the 
$\omega$-meson and the s-wave annihilation of pions at finite baryon 
density and temperature, are directly linked to the polarization operator, 
as indicated by the diagrams of Fig. 2.
\vspace{-0.1cm}
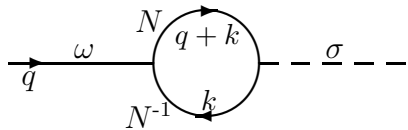
\begin{figure}[h]
\begin{picture}(150,70)(-195,0)
\thicklines
\put(-30,30){\line(1,0){55}}
\put(-20,30){\vector(1,0){4}}
\put(45,30){\circle{40}}
\put(44,50){\vector(1,0){5}}
\put(46,10){\vector(-1,0){5}}
\put(-5,32){\makebox(0,0)[bl]{$\omega$}}
\put(-25,27){\makebox(0,0)[tl]{$q$}}
\put(18,42){\makebox(0,0)[bl]{$N$}}
\put(14,15){\makebox(0,0)[tl]{$N^{\mbox{\scriptsize -1}}$}}
\put(33,45){\makebox(0,0)[tl]{\footnotesize $q+k$}}
\put(43,20){\makebox(0,0)[tl]{\footnotesize $k$}}
\put(65,30){\line(1,0){7}}
\put(77,30){\line(1,0){7}}
\put(89,30){\line(1,0){7}}
\put(101,30){\line(1,0){7}}
\put(113,30){\line(1,0){7}}
\put(90,32){\makebox(0,0)[bl]{$\sigma$}}
\end{picture}
\caption{The $\omega$$\sigma$ polarisation in nuclear matter.}
\end{figure}
\vspace{0.1cm}
\begin{figure}[h]
\begin{picture}(170,130)(-175,-75)
\put(-20,30){\makebox(0,0)[l]{a)}}
\put(15,30){\line(1,0){31}}
\put(19,32){\makebox(0,0)[bl]{$\omega$}}
\put(66,30){\circle{40}}
\put(41,45){\makebox(0,0)[bl]{$N$}}
\put(34,15){\makebox(0,0)[tl]{$N^{\mbox{\scriptsize -1}}$}}
\put(86,30){\line(1,0){7}}
\put(98,30){\line(1,0){7}}
\put(110,30){\line(1,0){7}}
\put(99,32){\makebox(0,0)[bl]{$\sigma$}}
\put(117,30){\line(1,3){10}}
\put(117,30){\line(1,-3){10}}
\put(121,50){\makebox(0,0)[br]{$\pi$}}
\put(121,10){\makebox(0,0)[tr]{$\pi$}}

\put(-20,-50){\makebox(0,0)[l]{b)}}

\put(15,-50){\line(-1,3){10}}
\put(15,-50){\line(-1,-3){10}}
\put(10,-30){\makebox(0,0)[bl]{$\pi$}}
\put(10,-70){\makebox(0,0)[tl]{$\pi$}}
\put(15,-50){\line(1,0){7}}
\put(27,-50){\line(1,0){7}}
\put(39,-50){\line(1,0){7}}
\put(28,-48){\makebox(0,0)[bl]{$\sigma$}}
\put(66,-50){\circle{40}}
\put(41,-35){\makebox(0,0)[bl]{$N$}}
\put(34,-65){\makebox(0,0)[tl]{$N^{\mbox{\scriptsize -1}}$}}
\put(86,-50){\line(1,0){20}}
\put(90,-48){\makebox(0,0)[bl]{$\omega$}}
\put(106,-50){\circle*{4}}
\put(106,-50){\line(1,0){20}}
\put(110,-47){\makebox(0,0)[bl]{$\gamma$}}
\put(126,-50){\line(1,3){10}}
\put(126,-50){\line(1,-3){10}}
\put(150,-30){\makebox(0,0)[br]{$e^+$}}
\put(150,-70){\makebox(0,0)[tr]{$e^-$}}
\end{picture}
\caption{In-medium $\omega$$\sigma$ mixing contributions to the
$\omega$-meson width (a) and to the s-wave $\pi\pi$ $\rightarrow$ $e^+e^-$
process (b).}
\end{figure}
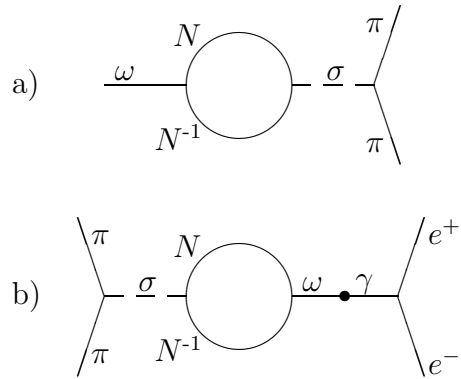

We calculate the polarization operator of Fig. 1 
in the relativistic mean-field approximation. 
Nuclear matter is treated as a system of noninteracting fermions
at finite density and temperature. We use free space 
meson fields. 
The $\sigma$- and $\omega$-meson fields 
couple to the scalar baryon density and to the
conserved baryon current respectively,
\begin{equation}
{\cal L}_{\sigma NN} = g_{\sigma} \overline \Psi \sigma \Psi
\end{equation}
and
\begin{equation}
{\cal L}_{\omega NN} = -g_{\omega} \overline \Psi \gamma _\mu \omega^\mu 
\Psi.
\end{equation}
The $\omega \sigma$ polarization at finite density and temperature is 
evaluated in the rest frame of nuclear matter. We generalize 
methods used previously \cite{Celenza,Henning-Friman,Lim} 
in calculations restricted to matter at zero temperature. 
The polarization tensor for the process 
shown in Fig. 1 is given by the expression,
\begin{equation}
-i \Pi^\mu =
-2 \, g_\sigma \, g_\omega \,    
\int {d^4 k \over (2\pi)^4} 
          \, Tr[\gamma^\mu iG(k+q)\, iG(k)], 
\end{equation}
where $iG(k)$ is the propagator of a nucleon in nuclear matter at 
finite temperature. The minus sign arises from the fermion loop and the factor
of 2 is the isospin degeneracy factor for symmetric nuclear matter. 
The nucleon propagator can be written as the sum of two terms,
\begin{equation}
 i G(k) =          
 iG_0(k) + iG_M(k), 
\end{equation}
the free propagator $iG_0(k)$ and a medium term $iG_M(k)$. At zero
temperature, the latter allows for the propagation of holes in the 
Fermi sea and restricts particle propagation to states above the 
Fermi surface.
\par
In the mean-field approximation, and neglecting the contribution of 
thermally excited antibaryons, the nucleon propagator 
reads
\begin{equation}
iG(k) = i(k \ \mkern-23mu\not \  +M_N^*) \, 
\Bigl\{ {1\over
k^{2} -M_N^{*2}+i\epsilon} +{i\pi \over E^*(k)} 
\delta \lbrack k^{0} - E^*(k) \rbrack \,
n_T (k)\Bigr \},
\end{equation}
where we have introduced the nucleon effective mass $M_N^*$,
the effective kinetic energy $E^*(k) = (\vec k \, 
^2 + M_N^{*2})^{1/2}$
and the nucleon thermal distribution function $n_T(k)$
\begin{equation}
n_T(k) = {1 \over {1+ exp \{\lbrack E^*(k)-\mu\rbrack /k_BT \} }}.
\end{equation}
In Eq. (6), $\mu$ is the baryon chemical potential and $k_B$ the Boltzmann constant.
At zero temperature, the thermal distribution function
$n_T(k)$ is replaced by the step function $\theta(k_F -|~\vec k~|)$.

Evaluating the trace, we have
\begin{eqnarray}
-i \Pi^\mu = 8\, M_N^* g_\sigma \, g_\omega     
\int {d^4 k \over (2\pi)^4} (2k+q)^\mu 
\Bigl\{ {1\over
k^{2} -M_N^{*2}+i\epsilon} +{i\pi \over E^*(k)}\, 
\delta \lbrack k^{0} - E^*(k) \rbrack \,
n_T (k)\Bigr \} \nonumber \\
\mkern 18 mu \Bigl\{ {1\over
(k+q)^2 -M_N^{*2}+i\epsilon} +{i\pi \over E^*(k+q)} 
\, \delta \lbrack k^{0}+q^{0} - E^*(k+q) \rbrack \,
n_T (k+q)\Bigr \}.
\end{eqnarray}
We need $\Pi^\mu$ only for time-like values of $q^{2}$.
The product of the propagators gives four terms. 
The first one is the vacuum polarization 
which is zero.
The product of the matter terms vanishes for time-like 
$q^{2}$ as the constraints from the $\delta$-functions, 
$\delta \lbrack k^{0} - E^*(k) \rbrack $ and
$\delta \lbrack k^{0}+q^{0} - E^*(k+q) \rbrack$, are in this case
incompatible. We are left with the two terms linear in $n_T$, which
are free of divergences.  
\par 
The polarization $\Pi^\mu$ satisfies
\begin{equation}
q_{\mu}\Pi^\mu = 0,
\end{equation}
as the $\omega$-field is coupled to a conserved current. 
We choose a coordinate system such that $\vec q$ is along
the z axis, i.e. $q_{\mu}=(q_{0},0,0,q_z)$. 
The $x$ and $y$ components of the polarization vanish in this   
frame due to the axial symmetry. Furthermore,
Eq. (8) implies 
\begin{equation}
q_{0}\Pi_0 = q_z\Pi_z,
\end{equation}
i.e. there is only one independent component in the polarization tensor. 
We calculate $\Pi^0$. After
integration over $k^{0}$ and  
a change of variable $\vec k \rightarrow \vec k + \vec q$ 
in the second term, we obtain
\begin{eqnarray}
\Pi^0 &=& 
 8 \, M_N^*  g_\sigma \, g_\omega \,    
\int {d^3 \vec k \over (2\pi)^3} \Bigl\{ 
{1 \over 2E^*(k)} 
\, {{2E^*(k)+q^{0}}\over
{E^*(k+q)^2-[E^*(k)+ q^{0}]^2 -i\epsilon}} \,
n_T (k) \nonumber \\
&& \; \; \; \; +\, {1 \over 2E^*(k)} 
\, {2E^*(k)-q^{0}\over
E^*(k-q)^2 -[E^*(k)- q^{0}]^2-i\epsilon} \,
n_T (k)\Bigr \} \\[3mm]
&\equiv& \Pi^0_1 + \Pi^0_2.
\end{eqnarray}

\noindent
Clearly, $\Pi^0$ is invariant separately under $q_0 \rightarrow -q_0$ and
$\vec q \rightarrow - \vec q$, while $\Pi_z$ is odd under those
transformations. It is then sufficient to consider $q_0 >0$.
In general, the polarization tensor in matter  
is a function of $\vec q$ as well as of $q_0$.
\par
The angular integrals in Eq. (10) are readily performed. We find
\par
\begin{eqnarray}
\Re e \, \Pi^0_1 &=& \mkern 5mu
 {M_N^* g_\sigma \, g_\omega\over 2 \pi^2 q_z}     
\int kdk \, {2E^*(k)+q^{0} \over E^*(k)}
\, \ln \, {{q^{2}+ 2E^*(k)q^{0} - 2kq_z} \over
{q^{2} +2E^*(k)q^{0}  
+ 2kq_z}} \, n_T (k), \\
\Im m \, \Pi^0_1 &=& \mkern 5mu 0.
\end{eqnarray}
\noindent
for $q^2>0$ and $q^0>0$.

For $\Pi^0_2$, there are two cases. For $q_0>0$ and 
$0<q^{2}< 4M_N^{*2}$, we have 
\begin{eqnarray}
\Re e \, \Pi^0_2 &=& \mkern 5mu
 {M_N^* g_\sigma \, g_\omega\over 2 \pi^2 q_z}     
\int kdk \, {2E^*(k)-q^{0} \over E^*(k)}
\, \ln \, {{q^{2} -2E^*(k)q^{0}- 2kq_z}
\over
{q^{2}- 2E^*(k)q^{0} + 2kq_z}}  
\, n_T (k), \\
\Im m \, \Pi^0_2 &=& \mkern 5mu 0.
\end{eqnarray}
\noindent
For $q^{2}> 4M_N^{*2}$, the real part reads  
\begin{equation}
\mkern -90mu \Re e \, \Pi^0_2= 
\mkern 5mu
{M_N^* g_\sigma \, g_\omega\over 2 \pi^2 q_z}     
\int kdk \, {2E^*(k)-q^{0} \over E^*(k)}
\, \ln \left | {{q^{2} -2E^*(k)q^{0}- 2kq_z}
\over
{q^{2}- 2E^*(k)q^{0} + 2kq_z}} \right | 
 n_T (k)
\end{equation}
\noindent
and the imaginary part is given by
\begin{eqnarray}
\Im m \, \Pi^0_2 &=& \mkern 5mu
{M_N^* g_\sigma \, g_\omega\over 2 \pi q_z}     
\int kdk \, {2E^*(k)-q^{0} \over E^*(k)}
\, \Theta \left[ {\left( {q_z \over 2} + {q^{0} \over 2} \sqrt {1-{4M_N^{*2} 
\over q^{2}}} \right) - k} \right] \nonumber \\
&& \mkern 200 mu\Theta \left[ k - \left( {q_z \over 2} 
- {q^{0} \over 2} \sqrt {1-{4M_N^{*2} 
\over q^{2}}} \right) \right]
 n_T (k)
\end{eqnarray}
\noindent
for $q^{2}< 2 q^{0}M_N^{*}$ and by
\begin{eqnarray}
\Im m \, \Pi^0_2 &=& \mkern 5mu
{M_N^* g_\sigma \, g_\omega\over 2 \pi q_z}     
\int kdk \, {2E^*(k)-q^{0} \over E^*(k)}
\, \Theta \left[ {\left( {q_z \over 2} + {q^{0} \over 2} \sqrt {1-{4M_N^{*2} 
\over q^{2}}} \right) - k} \right] \nonumber \\
&&\mkern 200 mu\Theta \left[ k + \left( {q_z \over 2} 
- {q^{0} \over 2} \sqrt {1-{4M_N^{*2} 
\over q^{2}}} \right) \right]
 n_T (k)
\end{eqnarray}
for $q^{2}> 2 q^{0}M_N^{*}$.  

The $ \omega \sigma$ mixing matrix elements depend on 
the $\sigma NN$ and $\omega NN$ coupling constants,
$g_\sigma$ and $g_\omega$, and on the nucleon effective mass, $M_N^*(\rho)$.
\par
We choose $g_\sigma$=12.78 from Ref. \cite{Durso}.
There is some uncertainty in this quantity linked to the
nature of the $\sigma$ field. The Walecka Model \cite{Walecka}
and the One-Boson-Exchange potentials \cite{Machleidt} favor somewhat smaller
values, on the order of $g_\sigma$=10. 
\par
For the $\omega NN$ coupling constant,
we adopt $g_\omega$=9 as required by universality
\cite{Meissner}. We expect this value to be
the most adequate one for describing the coupling of physical
$\omega$-mesons to nucleons.
In the Walecka Model, the preferred value
is $g_\omega$=11.7 \cite{Walecka} while 
an even larger coupling constant $g_\omega \simeq $17 is used 
in One-Boson-Exchange potentials \cite{Machleidt}.
It is likely that various other contributions to the repulsive
part of the nucleon-nucleon interaction are subsumed in 
the effective $\omega$ degree of freedom in these approaches.
\par
To calculate the diagrams 
displayed in Fig. 2, we shall evaluate
the polarization at $q^2=m_\omega^2$. 
We note that, for $q^2=m_\omega^2$, a form factor of the form
[($\Lambda^2 - m_\omega^2)/(\Lambda^2 - q^2)]^n$ at the $\omega NN$ vertex
is strictly equal to unity. Similarly, a form factor of the same form at the
$\sigma NN$ vertex, but with $m_\omega$ replaced by $m_\sigma$, is very close 
to unity for the same value of $q^2$ and for $m_\sigma$=0.84 GeV (see below). 
Consequently, such form factors can be neglected in our calculation.
\par
The density dependence of the nucleon effective mass is determined within
the nonlinear field theoretical approach of Ref. \cite{Boguta}.  
The Lagrangian of this model is obtained by adding 
nonlinear terms of the form,
\begin{equation}
{\cal L}^{NL} = -{a M_N \over 3} g_\sigma^3 \sigma^3
-{b \over 4} g_\sigma^4 \sigma^4,
\end{equation}
\noindent
to the Walecka scalar-vector model \cite{Walecka}. 
The parameters of the Lagrangian are fitted to give a reasonable
description of the properties of nuclear matter. We impose as 
contraints the saturation density ($\rho_0
=0.17$ fm$^{-3}$), the binding energy per nucleon (E/A=-16 MeV) 
and the compression modulus (K=300 MeV). We note that this model
allows for values of K  
much closer to the empirical value than the very large 
incompressibility (K$\simeq$550 MeV) obtained in the original Walecka model 
\cite{Walecka}. Furthermore, we fix the nucleon  
effective mass at saturation density to be 
$M^*_N(\rho_0)=0.7 \, M_N$, 
a value appropriate for particle-hole excitations
involving momenta far above the Fermi surface \cite{Mahaux}.
However, since our results are very sensitive to the value of the
nucleon effective mass,
we also give some results obtained with $M^*_N(\rho_0)=0.8 \, M_N$.  
The four parameters of the model are $C_{\sigma}^2=g_\sigma^2(M_N/m_\sigma)^2$,
$C_\omega^2=g_\omega^2(M_N/m_\omega)^2$, $A=a M_N g_\sigma^3$
and $B=b g_\sigma^4$. Their numerical values are $C_\sigma^2=243.22$,
$C_\omega^2=145.77$, $a= 0.321 \, 10^{-2}$ and $b=-0.129 \, 10^{-2}$
for $M^*_N(\rho_0)=0.7 \, M_N$ and $C_\sigma^2=173.35$,
$C_\omega^2=85.82$, $a= 0.256 \, 10^{-2}$ and $b= 0.309 \, 10^{-1}$
for $M^*_N(\rho_0)=0.8 \, M_N$. 
\par
\vskip 0.4 truecm
\noindent
{\bf 3. $\pi\pi$ scattering phase shifts in the s-wave}\par
\vskip 0.4 truecm
\noindent
We present in this section a model for $\pi \pi$ scattering in the s-wave 
and fix the $\sigma \pi \pi$ vertex (coupling constant and form 
factor in the time-like region) as well as the $\sigma$-meson mass,
needed to calculate the processes illustrated in Fig. 2.

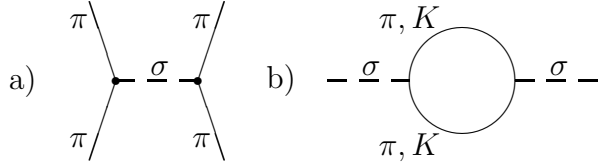
\begin{figure}
\begin{picture}(150,70)(-140,0)
\put(-10,30){\makebox(0,0)[r]{a)}}
\put(20,30){\line(-1,3){10}}
\put(20,30){\line(-1,-3){10}}
\put(6,50){\makebox(0,0)[b]{$\pi$}}
\put(6,10){\makebox(0,0)[t]{$\pi$}}
\put(20,30){\circle*3}
\put(20,30){\line(1,0){7}}
\put(32,30){\line(1,0){7}}
\put(44,30){\line(1,0){7}}
\put(33,32){\makebox(0,0)[bl]{$\sigma$}}
\put(51,30){\circle*{3}}
\put(51,30){\line(1,3){10}}
\put(51,30){\line(1,-3){10}}
\put(56,50){\makebox(0,0)[br]{$\pi$}}
\put(56,10){\makebox(0,0)[tr]{$\pi$}}

\put(88,30){\makebox(0,0)[r]{b)}}

\put(100,30){\line(1,0){7}}
\put(112,30){\line(1,0){7}}
\put(124,30){\line(1,0){7}}
\put(113,32){\makebox(0,0)[bl]{$\sigma$}}
\put(151,30){\circle{40}}
\put(131,50){\makebox(0,0)[b]{$\pi,K$}}
\put(131,10){\makebox(0,0)[t]{$\pi,K$}}
\put(171,30){\line(1,0){7}}
\put(183,30){\line(1,0){7}}
\put(195,30){\line(1,0){7}}
\put(184,32){\makebox(0,0)[bl]{$\sigma$}}
\end{picture}
\caption{$\pi \pi$ s-wave scattering (a) and $\sigma$ self-energy
(b).}
\end{figure}

We assume that s-wave $\pi\pi$ scattering is dominated by the $\sigma$-meson 
in the s-channel
and we compute the $\sigma$ self-energy to the one-loop order, including
both pion and kaon loops, as shown in Fig. 3.
We remark that this approach is very different from meson-exchange models 
in which a potential for $\pi \pi$ scattering is generated by summing 
the lowest-lying s-, t- and u-channel exchange diagrams \cite{Janssen}. 

In resonant scattering, the phase shift is directly related to the
resonance propagator. In our model of s-wave $\pi \pi$ scattering,
the phase shift $\delta _0$ below the $K \bar K$ threshold can be
expressed as a function of the real and 
imaginary parts of the $\sigma$-meson propagator $G_\sigma$,
\begin{equation}
\tan(\delta_0(q)) = {\Im m \, G_\sigma(q) \over \Re e \, G_\sigma(q)} =
- {\Im m \, G_\sigma^{-1}(q) \over \Re e \, G_\sigma^{-1}(q)},
\end{equation}
\noindent
where $G_\sigma$ is defined by
\begin{equation}
G_\sigma(q)={1 \over {q^2-m{^0_\sigma}^{2}-\Sigma(q)}}.
\end{equation}
In Eq. (21), $m^0_\sigma$ is the bare $\sigma$-meson mass and $\Sigma(q)$
the self-energy associated with the $\pi$
and K loops of Fig. 3b.
\par
In order to calculate $\Sigma(q)$, we assume the following forms
of the interaction Lagrangians
\begin{equation}
{\cal L}_{\sigma \pi \pi} 
= {1 \over 2}\, g_{\sigma \pi \pi} \, m_\pi \,\vec \pi \mkern 3 mu \vec \pi
\sigma 
\end{equation}
\noindent
and
\begin{equation}
{\cal L}_{\sigma K K} 
=  g_{\sigma K K} \, m_K \, (\overline K^0 K^0 \sigma + K^+ K^-
\sigma) .
\end{equation}
\noindent
We use extended $\sigma\pi\pi$ and $\sigma KK$ vertices characterized by the 
same functional form and cutoff,
\begin{equation}
F(q)= {\Lambda^2 \over {\Lambda^2-q^2}}.
\end{equation}
\noindent
The form factor effectively accounts for higher order processes,
not included in the lowest order approximation.
\par
The self-energy $\Sigma(q)$ is given by
\begin{eqnarray}
-i \Sigma(q) \, &=& \,
 {3 \over 2}\,  g_{\sigma\pi \pi}^2 \, m_\pi^2 \, F^2(q^2)\,   
\int {d^4 k_1 \over (2\pi)^4} 
\, {1 \over {k_1^2 -m_\pi^2 +i\epsilon}} 
\, {1 \over {(k_1-q)^2 -m_\pi^2 +i\epsilon}} \nonumber \\
&&\mkern 30mu+\, 
 2\,  g_{\sigma KK}^2 \, m_K^2 \, F^2(q^2)\,       
\int {d^4 k_2 \over (2\pi)^4} 
\, {1 \over {k_2^2 -m_K^2 +i\epsilon}} 
\, {1 \over {(k_2-q)^2 -m_K^2 +i\epsilon}},
\end{eqnarray}
where the coefficient 3/2 for the pion loop comes from the 
the 3 isospin states and the permutation symmetry factor
and the coefficient 2 for the kaon loop accounts for the neutral and 
charged kaon contributions. We are interested in fitting 
the s-wave $\pi \pi$ phase shifts for values of $\sqrt s$
of the order of the $\omega$-meson mass, i.e. typically
between $\sqrt s=2 m_\pi$ and  $\sqrt s = 2m_K$. 
In this range, $4 m_\pi^2 < q^2 < 4 m_K^2$, 
the contribution of the pion loop
to $\Sigma(q)$ has both real and imaginary parts  
while the contribution of the kaon loop is real.
\par
The pion and kaon loops are logarithmically divergent. We regularize the
correspon\-ding integrals using the Pauli-Villars method \cite{Pauli-Villars}.
We couple for 
each loop the $\sigma$-meson to an additional pseudoscalar field
with a very large mass and multiply the regularizing integrals
by constants chosen so as to remove the divergences. This procedure is 
discussed in detail for the calculation of the vacuum polarization 
in quantum electrodynamics in Ref. \cite{Itzykson-Zuber}.
\par
For $4 m_\pi^2 < q^2 < 4 m_K^2$, 
we find 
\begin{eqnarray}
\Re e \, \Sigma(q) &=&
- {3\over {16 \pi^2}}\,g_{\sigma\pi \pi}^2 \, m_\pi^2 \,
F^2(q^2)
\, \left\lbrace \ln {m_\Lambda \over m_\pi} 
+{1 \over 2}
\left ({{q^2-4 m_\pi^2}\over {q^2}}\right )^{1/2}
\ln \left[{{q-(q^2-4m_\pi^2)^{1/2}}\over {q+(q^2-4m_\pi^2)^{1/2}}}
\right] \right. \nonumber \\
&& \mkern 50mu \left. -{1 \over 2}
\left ({{4 m_\Lambda^2-q^2}\over {q^2}}\right )^{1/2} 
\left[2\arctan \left ({{4 m_\Lambda^2-q^2}\over {q^2}}\right )^{1/2} 
- \pi \right] \right\rbrace \nonumber \\
&& 
- {1\over {4 \pi^2}}\,g_{\sigma KK}^2 \, m_K^2 \, F^2(q^2)
\left\lbrace \ln {m_{\Lambda'} \over m_K} \right.\nonumber \\
&&\mkern 50mu \left.
+{1 \over 2}
\left ({{4 m_K^2-q^2}\over {q^2}}\right )^{1/2} 
\left[2\arctan \left ({{4 m_K^2-q^2}\over {q^2}}\right )^{1/2} 
- \pi \right] \right. \nonumber \\
&& \mkern 50mu \left. -{1 \over 2}
\left ({{4 m_{\Lambda'}^2-q^2}\over {q^2}}\right )^{1/2} 
\left[2\arctan \left ({{4 m_{\Lambda'}^2-q^2}\over {q^2}}\right )^{1/2} 
- \pi \right] \right\rbrace 
\end{eqnarray}
and
\smallskip
\begin{equation} 
\Im m \, \Sigma(q)=
- {3\over {32 \pi}}\,g_{\sigma\pi \pi}^2 \, m_\pi^2 \,
F^2(q^2)
\, \left ({{q^2-4 m_\pi^2}\over {q^2}} \right)^{1/2},
\end{equation}
\noindent
where $m_\Lambda$ and $m_{\Lambda'}$ are the Pauli-Villars masses
and $q=\sqrt {q^2}$.
The value of the physical mass of the $\sigma$-meson in our model is fixed by 
the condition
\begin{equation}
\Re e \, G_\sigma^{-1}(q^2= m_\sigma^2)=0. 
\end{equation}
This relation provides the functional dependence of $m{^0_\sigma}^2$ 
on
$m_\sigma^2$. Inserting this value of $m{^0_\sigma}^2$ in the expression for
$G_\sigma(q)$, we obtain
\begin{equation}
\Re e \, G_\sigma^{-1}(q^2)= q^2-m_\sigma^2-(\Sigma(q^2)-
\Sigma (m_\sigma^2)). 
\end{equation}
For $m_\Lambda$ and $m_{\Lambda'}$
approaching infinity, the terms involving
the Pauli-Villars masses vanish and the physical 
contribution to $\Re e \, G_\sigma^{-1}(q^2)$ is given
by the remaining terms,
\begin{eqnarray}
 \Re e \, G_\sigma^{-1}(q^2) &=& q^2-m_\sigma^2 \nonumber \\
&& \mkern -50mu - {3\over {16 \pi^2}}\,g_{\sigma\pi \pi}^2 \, m_\pi^2 \left\lbrace 
F^2(m_\sigma^2)
\left ({{m_\sigma^2-4 m_\pi^2}\over {4 m_\sigma^2}}\right )^{1/2}
\ln \left[{{m_\sigma-(m_\sigma^2-4m_\pi^2)^{1/2}}\over {m_\sigma+(m_\sigma^2
-4m_\pi^2)^{1/2}}}
\right] \right. \nonumber \\
&& \; \; \; \left.
-F^2(q^2)
\left ({{q^2-4 m_\pi^2}\over {4 q^2}}\right )^{1/2}
\ln \left[{{q-(q^2-4m_\pi^2)^{1/2}}\over {q+(q^2
-4m_\pi^2)^{1/2}}}
\right] \right\rbrace \nonumber \\
&& \mkern -50mu
- {1\over {4 \pi^2}}\,g_{\sigma KK}^2 \, m_K^2 \, 
\left\lbrace F^2(m_\sigma^2)\, 
\left ({{4 m_K^2-m_\sigma^2}\over {4m_\sigma^2}}\right )^{1/2} 
\left[2\arctan \left ({{4 m_K^2-m_\sigma^2}\over {m_\sigma^2}}\right )^{1/2} 
- \pi \right] \right. \nonumber \\
&&  \; \; \; \left.- 
F^2(q^2) \,
\left ({{4 m_K^2-q^2}\over {4q^2}}\right )^{1/2} 
\left[2\arctan \left ({{4 m_K^2-q^2}\over {q^2}}\right )^{1/2} 
- \pi \right]  
\right\rbrace 
\end{eqnarray}
\noindent
\begin{figure}[t]
 \begin{center}
\hspace*{-10mm}  \mbox{\epsfig{file=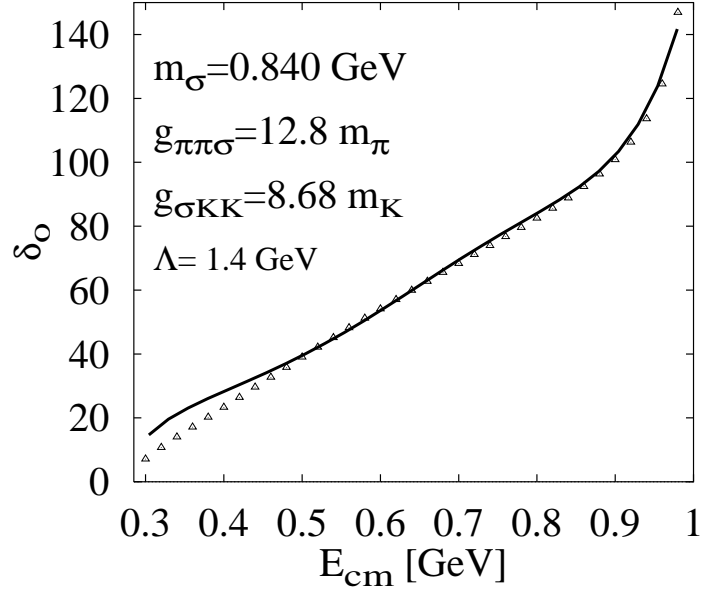,height=8.0cm}}
 \end{center}
\vspace{-3mm}
\caption{%
Fit to the $\pi \pi$ s-wave scattering phase shift.
}
\end{figure}
and
\begin{equation}
\Im m \, G_\sigma^{-1}(q^2)= 
{3\over {32 \pi}}\,g_{\sigma\pi \pi}^2 \, m_\pi^2 \,
F^2(q^2)
\, \left ({{q^2-4 m_\pi^2}\over {q^2}} \right)^{1/2}.
\end{equation}
Using Eq. (~20), we can fit the s-wave $\pi \pi$
phase shift $\delta_0$. The result, displayed in Fig. 4,
is obtained with $m_\sigma$=0.84 GeV, $g_{\sigma \pi \pi}$=12.8,  
$g_{\sigma KK}$=8.68 and $\Lambda$=1.4 GeV. Considering the 
complexity of the $\sigma$ degree of freedom
and the simplicity of the model, this fit is remarkably
good. It is particularly precise in the region of interest for our problem,
i.e. around $\sqrt s \simeq m_\omega$.     

\noindent
{\bf 4. Numerical results}\par
\vskip 0.4 truecm
\noindent
{\bf 4.1. In-medium broadening of $\omega$-mesons of finite momenta}\par
\vskip 0.4 truecm
\noindent
In this subsection, we calculate the process illustrated in Fig. 2a.
We consider an $\omega$-meson on its mass-shell ($q^2=m_\omega^2,
q_z \not = 0$) which converts into a $\sigma$-meson through a 
particle-hole excitation and decays into two pions of momenta $k_1$
and $k_2$ respectively. \par
In the nuclear matter rest frame, the partial decay width for this
process is given by \cite{Itzykson-Zuber} 
\begin{equation}
\Gamma_{\omega \rightarrow \pi \pi} = 
\mkern 5mu
{1 \over 24 \pi^2 q^0} {3 \over 2}     
\int {d^3k_1 \over 2k_1^0}{d^3k_2 \over 2k_2^0}
 \, \sum_{spins}
\, \left | M \right |^2 \delta ^4 (q-k_1-k_2),
\end{equation}
\noindent
where M is the invariant transition matrix element and the
factor 3/2 accounts for the sum over isospin 
and the symmetry of the final state.
\par
Using the interaction Lagrangian (22) and the results 
of Section 2, a straightforward calculation yields
\begin{equation}
 \Gamma_{\omega \rightarrow \pi \pi} = 
{(g_{\sigma \pi \pi} m_\pi)^2 \over 4 \pi}\, F^2(q^2=m_\omega^2)\,
 {k_\pi \over 4 q_z^2}\,     
{m_\omega \over q^0} \,
\mid G_\sigma(q^2=m_\omega^2)\mid^2 
\, \left[ \Pi^0(q^2=m_\omega^2, q_z) \right]^2,
\end{equation}
\noindent
where the form factor F is defined by Eq. (24), $k_\pi=
\sqrt{m_\omega^2/4-m_\pi^2}$ is the pion 3-momentum in the
$\omega$ rest frame, $G_\sigma$ and $\Pi^0$ are the $\sigma$
propagator [Eqs. (30) and (31)] and the $ \omega \sigma$
polarization [Eqs. 12-18] respectively.

In Figs. 5-7, we show  
$\Gamma_{\omega \rightarrow \pi \pi}$ 
as a function of density, $\omega$-meson 3-momentum and temperature.   
In each figure, we indicate also the free $\omega$ width which
should be added to the matter contribution (33).
\begin{figure} 
 \begin{center}
\hspace*{-1cm}  \mbox{\epsfig{file=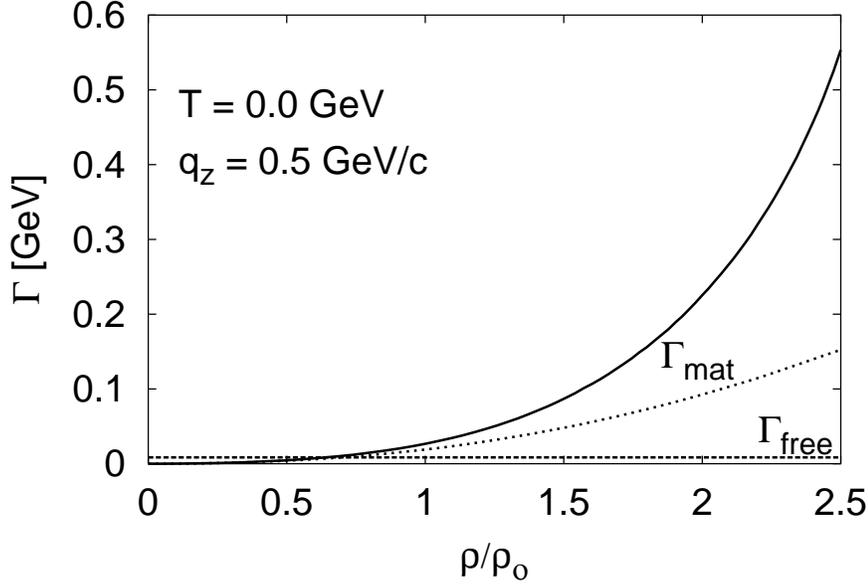,height=8.5cm}}
\vspace{-5mm}
 \end{center}
\caption{%
Density dependence
of $\Gamma_{\omega \rightarrow \pi \pi}$ 
(denoted $\Gamma_{mat}$)
at T=0 and for an $\omega$ 3-momentum $q_z$= 0.5 GeV/c. 
The full curve is the result obtained with 
$M^*_N(\rho_0)=0.7 \, M_N$; the dotted line shows the effect 
of using $M^*_N(\rho_0)=0.8 \, M_N$ instead.
$\Gamma_{free}$ is the $\omega$ width in free space
(dashed line).
}
\label{densdepwidth}
\end{figure}
\begin{figure}[h] 
 \begin{center}
\hspace*{-1cm}   \mbox{\epsfig{file=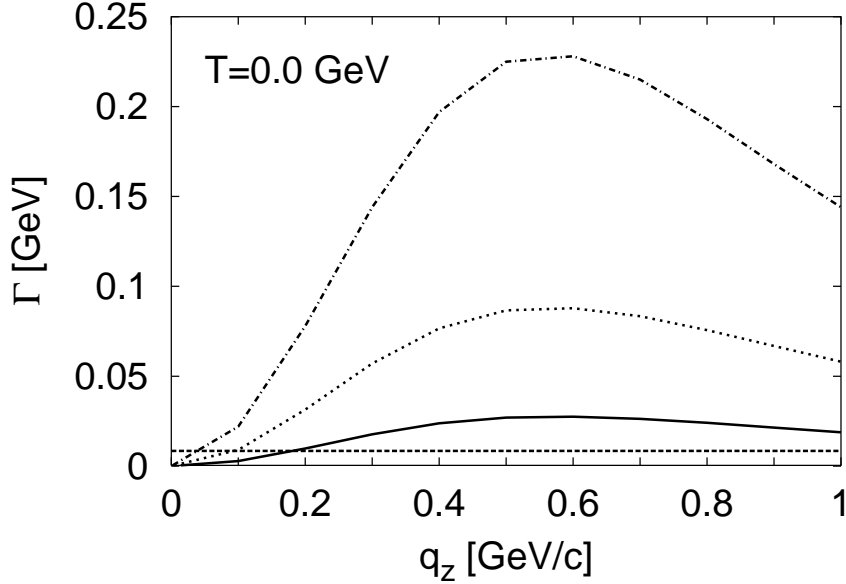,height=8.5cm}}
 \end{center}
\vspace{-5mm}
\caption{%
Momentum  dependence
of $\Gamma_{\omega \rightarrow \pi \pi}$ 
at T=0 and for 3 values of the matter 
density $\rho/\rho_0$ = 1 (full line), 1.5 (dotted line) and 2
(dot-dashed line). The free width is indicated by a dashed line.
}
\label{momdepwidth}
\end{figure}

We discuss first the $\omega$ broadening due to the $\omega
\sigma$ polarization in nuclear matter at zero temperature (Fig. 5).
The most striking result is that $\Gamma_{\omega \rightarrow \pi \pi}$ 
is very large. At $\rho=\rho_0$, $\Gamma_{\omega \rightarrow \pi \pi}$ 
is 27 MeV for $q_z$=0.5 GeV/c, i.e. in the momentum range where 
the broadening is most pronounced.
The reason for this is twofold. On the one hand, as already
noted in Ref. \cite{Celenza}, the $ \omega \sigma$ polarization is 
large. On the 
other hand, the $\sigma$ propagator taken at $q^2=m_\omega^2$
enhances the effect. Its real
part is indeed very small since our value for $m_\sigma$ 
is close to $m_\omega$ [$m_\omega^2-m_\sigma^2=-0.094$ GeV$^2$].
Furthermore, $\Gamma_{\omega \rightarrow \pi \pi}$ 
is also very sensitive to the nucleon effective mass. We 
illustrate this effect in Fig. 5 by showing the values of
$\Gamma_{\omega \rightarrow \pi \pi}$ obtained for 
$M^*_N(\rho_0)=0.8 \, M_N$. 
With this choice for $M^*_N$, $\Gamma_{\omega \rightarrow \pi \pi}$ 
is reduced by about 8 MeV at $\rho=\rho_0$. 
The sensitivity of $\Gamma_{\omega \rightarrow \pi \pi}$ to the density 
dependence of the nucleon effective mass is very strong at larger densities,
where our estimate is therefore much more uncertain than at
$\rho=\rho_0$.  
We note that, in the range of densities considered in Fig. 5,
the $\omega \sigma$ polarization is real because
$q^2=m_\omega^2$ is always $<4 M_N^{*2}$. Hence, the in-medium 
width is due solely to the two-pion final state.
As expected, $\Gamma_{\omega \rightarrow \pi \pi}$ 
exhibits a strong dependence on
the baryon density.

\par
The 3-momentum dependence of $\Gamma_{\omega \rightarrow \pi \pi}$,
displayed in Fig. 6 at T=0 for $\rho/\rho_0$=1, 1.5 and 2,
illustrates the kinematic conditions under which the broadening
of the $\omega$-meson due to the $\omega \sigma$ polarization is the largest.
At $q_z = 0$, where the $\omega \sigma$ mixing is not allowed, 
$\Gamma_{\omega \rightarrow \pi \pi}$ vanishes.
As a consequence of the kinematic factors in Eq. (33), the initial 
growth of $\Gamma_{\omega \rightarrow \pi \pi}$ with $q_z$ 
saturates, and is reversed for $q_z >$ 0.6 GeV/c.

\begin{figure}[t] \begin{center}
\hspace*{-1cm}  \mbox{\epsfig{file=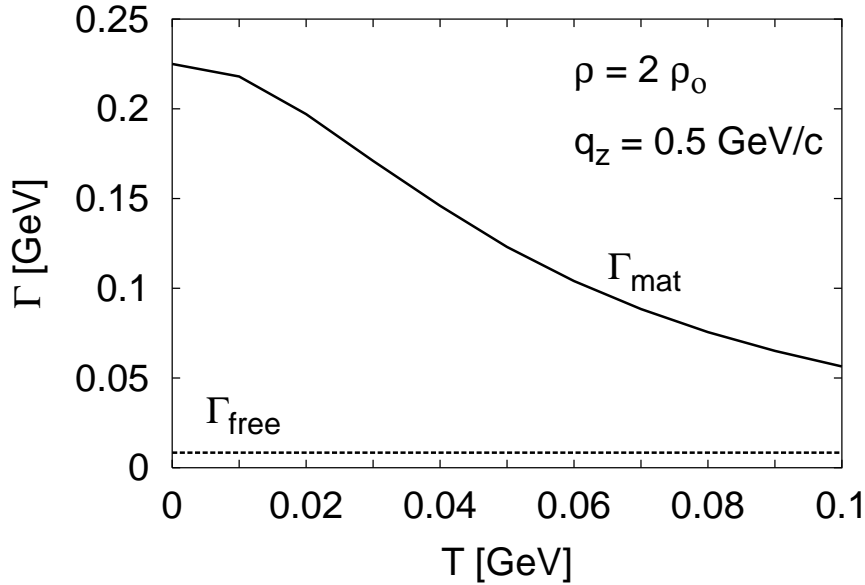,height=8.5cm}}
 \end{center}
\vspace{-5mm}
\caption{%
Temperature dependence
of $\Gamma_{\omega \rightarrow \pi \pi}$ 
(denoted $\Gamma_{mat}$)
at $q_z=0.5$ GeV/c and for $\rho/\rho_0$ = 2 (full line).
 $\Gamma_{free}$ is the $\omega$ width in free space
(dashed line).
}
\label{tempdepwidth}
\end{figure}

\noindent
In Fig. 7,
we show the temperature dependence of $\Gamma_{\omega \rightarrow \pi \pi}$
for $\rho/\rho_0$=2 and $q_z=0.5$ GeV/c. 
The in-medium $\omega$ width due to $\omega \sigma$ mixing decreases with 
increasing temperature.
This effect can be traced back to the temperature dependence of the nucleon
effective mass, which increases with increasing temperature in Walecka
models. 
For the temperatures
expected to prevail in relativistic heavy ions collisions 
(E/A $\simeq$ 1-2 GeV), $\Gamma_{\omega \rightarrow \pi \pi}$
is decreased by at least a factor of two compared to the T=0 value 
at the same density. 
We emphasize that this behavior is very much model dependent.
The increase of the nucleon effective mass with increasing temperature
at finite density (T $<$ 100 MeV) is a particular feature of the 
Walecka model. In models where the nucleon effective mass 
decreases with increasing temperature, $\Gamma_{\omega \rightarrow \pi \pi}$
would increase with increasing temperature.
\par

{\bf 4.2. S-wave $\pi \pi$ annihilation into $e^+e^-$ pairs
}\par

\vskip 0.4 truecm
\noindent
We discuss now the process illustrated in Fig. 2b and study
its importance in the production of $e^+e^-$ pairs in
relativistic heavy ion collisions.
\par
We calculate the s-wave annihilation of two pions ($\pi^+\pi^-$
or $\pi^0\pi^0$) of momenta $p_1$ and $p_2$ such that 
$(p_1+p_2)^2=m_\omega^2$. They annihilate in the $\sigma$ channel
which, through particle-hole excitations, mixes with the
$\omega$-meson. The $\omega$ can decay into an electron and
a positron of momenta $k_1$ and $k_2$.  
The diagram of Fig. 2b is evaluated in the rest frame 
of the nuclear medium (which we identify with the interaction region
in central heavy ion collisions). 
For each initial state ($\pi^+\pi^-$, $\pi^-\pi^+$ and $\pi^0\pi^0$),
the annihilation cross section is given by  
\begin{equation}
\label{eq:cross}
 \sigma_{\pi \pi \rightarrow e^+ e^-}^{s-wave} = 
\mkern 5mu
{1 \over 8 \pi^2 m_\omega} {m_e^2 \over \sqrt {m_\omega^2 - 4 m_\pi^2}} \,
\int {d^3k_1 \over k_1^0}{d^3k_2 \over k_2^0}
 \, \sum_{spins}
\, \left | M \right |^2 \delta ^4 (p_1+p_2-k_1-k_2).
\end{equation}
\noindent
The calculation of this cross section involves the same quantities as 
the partial decay width
of Eq. (33). We use 
the additional assumption of vector meson dominance for 
the electromagnetic current \cite{Kroll}. The conversion of an 
$\omega$-meson into a massive photon is characterized by the coupling
constant
\begin{equation}
f_\omega = {e m_\omega^2 \over 2 g_\omega}.
\end{equation}
\noindent
with $g_\omega$ defined as in Eq. (2).
Neglecting the electron mass compared to the electron momentum, we
obtain for the annihilation cross section  
\begin{eqnarray}
 \sigma_{\pi \pi \rightarrow e^+ e^-}^{s-wave} &=& 
\mkern 5mu
{\pi \over {3 q_z^2}}
{1 \over \sqrt {q^2 - 4 m_\pi^2}} 
\, \left ({\alpha \over g_\omega} \right)^{2}
\,(g_{\sigma \pi \pi} m_\pi)^2 \, {m_\omega^4 \over 
{(q^2-m_\omega^2)^2 + m_\omega^2 \Gamma_\omega^{tot\, 2}}}\,
F^2(q^2) \nonumber \\[2mm]
&&\mkern 100mu \times
\mid G_\sigma(q^2)\mid^2\, 
\left[ \Pi^0(q^2, q_z) \right]^2.
\end{eqnarray}
\par
\noindent
The width $\Gamma_\omega^{tot}$ entering in Eq. (36) is the sum
of $\Gamma_{\omega \rightarrow \pi \pi} $ given by Eq. (33)
and of the $\omega$ width in free space, $\Gamma_\omega^{free}$.
This expression includes therefore the in-medium broadening of
the $\omega$-meson discussed in 4.1. 
\par
To evaluate the importance of such a process in heavy ion collisions,
we compare the cross section for $\pi^+\pi^-$ annilation in s-wave 
[Eq.(36) for charged pions] to the corresponding cross section 
for $\pi^+\pi^-$ annilation in p-wave.
The latter reaction is known to be the dominant source of 
$e^+ e^-$ pairs in relativistic heavy ion collisions
(1$<\,$E/A$\,<$2 GeV) 
in the invariant mass region around the 
$\rho$- and $\omega$-meson mass, 0.6$<\,$m$_{e^+e^-}<$0.8 GeV
\cite{Brat96,Wolf95,Wolf93,Wolf90}. Its cross section in the 
$\rho$-dominance model is given by
\cite{Gale87}
\begin{equation}
 \sigma_{\pi^+ \pi^- \rightarrow e^+ e^-}^{p-wave} =
\mkern 5mu
{4 \pi \over {3}}
\, \left ({\alpha^2 \over q^2} \right)
{\sqrt {{q^2 - 4 m_\pi^2} \over q^2}} 
\, {m_\rho^4 \over 
{(q^2-m_\rho^2)^2 + m_\rho^2 \Gamma_\rho^{2}}}.
\end{equation}
\par
We calculate $\sigma_{\pi^+ \pi^- \rightarrow e^+ e^-}^{s-wave}$ and  
$\sigma_{\pi^+ \pi^- \rightarrow e^+ e^-}^{p-wave}$ at $\rho=2\, \rho_0$
and T=50 MeV, i.e. under conditions which characterize the interaction
region in central heavy ion collisions at relativistic energies. 
The cross section for $\pi^+\pi^-$ annilation in s-wave 
will depend not only on q$^2$ but also on q$_z$. We compute it for
q$_z$=0.25 GeV/c and q$_z$=0.5 GeV/c, to see the effect of the 
kinematics of the $\omega$-meson on the s-wave $\pi^+\pi^-$ annilation
cross section. For a proper comparison with the p-wave  
$\pi^+\pi^-$ annilation cross section, we would need a consistent
calculation of the $\rho$-meson width at $\rho=2\, \rho_0$ and T=50 MeV.
In the absence of such calculation and as the $\rho$-meson is expected
to broaden substantially at high density \cite{Klingl,rho}, we use both 
the free width
($\Gamma_\rho^{free}$) and twice the free width in the evaluation
of Eq. (37).
The result is shown in Fig. 8. It is clear that the total cross 
section for $\pi^+ \pi^-$ annihilation
in p-wave is larger than the total cross section for $\pi^+ \pi^-$ 
annihilation in s-wave in all cases. 
\noindent
\begin{figure} [h]
 \begin{center}
\hspace{-1cm}
  \mbox{\epsfig{file=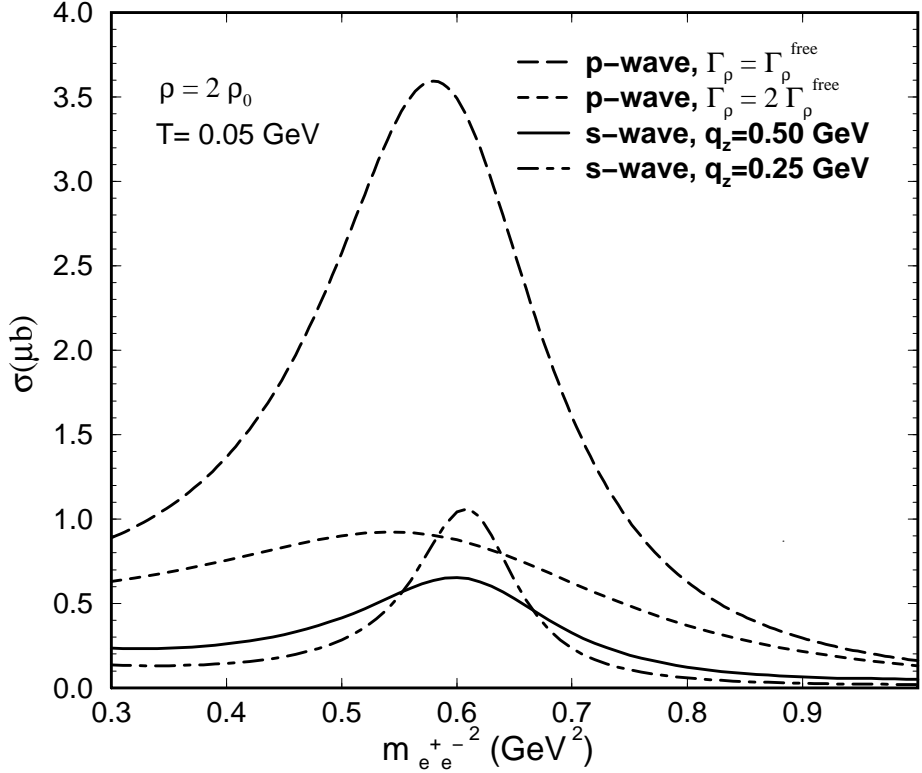,height=12cm}}
\end{center}
\caption{%
P-wave and s-wave $\pi^+ \pi^- $ annihilation cross sections into $e^+
e^-$ pairs. The long- and short-dashed lines show
the p-wave annihilation cross section [Eq. (37)] calculated with 
$\Gamma_\rho\,=\,\Gamma_\rho^{free}$ and
$\Gamma_\rho\,=\,2\Gamma_\rho^{free}$
respectively. The full and dot-dashed lines show the s-wave
annihilation cross section [Eq. (34)] for
 q$_z$=0.5 GeV/c and q$_z$=0.25 GeV/c.
}
\label{annihil}

\end{figure}

To observe the momentum-dependent broadening of the $\omega$-meson
discussed in Subsection 4.1, kinematic cuts in 
the momentum of $e^+ e^-$ pairs produced in 
relativistic heavy ion collisions would be most helpful.
At low momenta, the in-medium $\omega$-meson width will be smaller
and more likely to produce a structure in the $e^+ e^-$ spectrum,
in particular if the $\rho$-meson is very broad. However, this
structure
is expected to reflect the direct production of $\omega$-mesons 
in pion-nucleon and nucleon-nucleon collisions
\cite{Brat96,Wolf95,Wolf93,Wolf90}
rather than the s-wave
$\pi^+ \pi^-$ annihilation into $e^+ e^-$ pairs. 
Consequently, we do not anticipate that the observation
of the latter process will be possible.

\newpage

{\bf 5. Conclusion}\par
\vskip 0.4 truecm
\noindent
In this paper, we explore two consequences of a large
nondiagonal $\omega \sigma$ polarization in matter at zero and 
finite temperature. Our most important result is a substantial
broadening
 of $\omega$-mesons
moving with res\-pect to a nuclear medium.
The additional width, arising from the conversion of $\omega$-mesons 
into $\sigma$-mesons through particle-hole excitations and the 
subsequent decay of the $\sigma$ into two pions, can be much 
larger than the free width.
At T=0, this in-medium width increases with density and with the 
3-momentum of the $\omega$-meson for $q \lapp 0.5$ GeV/c.  
The effect decreases with
increasing temperature in the Walecka Model.
Our results are therefore most relevant for the studies of the 
propagation of
$\omega$-mesons in nuclei. In particular, the in-medium broadening of the 
$\omega$ meson should have observable consequences in experiments
where
 $\omega$-mesons are produced in 
heavy nuclear targets and observed in the $e^+e^-$ decay channel 
\cite{CEBAF,HADES}. As already emphasized, collisional broadening
from the $\omega \, N \rightarrow \pi \, N$ and  $\omega \, N 
\rightarrow \rho \, N$ reactions is expected to lead to a substantial 
further increase of the $\omega$ width in matter. All these processes
should be included consistently in a calculation of the $\omega$
propagator at finite baryon density. \par
Our study of the inverse process, the s-wave $\pi\pi$ annihilation
producing an $\omega$-meson which can decay into an $e^+e^-$ pair, 
shows that 
its contribution to dilepton spectra in relativistic heavy-ion collisions
is much smaller than that of p-wave $\pi^+\pi^-$ annihilation
in the $\rho$-channel. Consequently, we do not expect observable effects 
of this process in the 
total dilepton mass spectrum. 
However, the momentum dependent 
broadening of the $\omega$ meson may lead to a detectable effect also in 
heavy-ion collisions, once momentum cuts are applied.

\vskip 0.4truecm
{\bf Acknowledgements}\par
\vskip 0.4 truecm
\noindent
Two of us (M. S. and Gy. W.) gratefully acknowledge
the hospitality of GSI
where much of this work was done. Gy. W. was supported in part
by the Hungarian Research Foundation OTKA (grants T022931 and
T016594) and by the french MAE/CEA 1996 Convention.

\end{document}